\documentclass[%
 reprint,
 amsmath,amssymb,
 aps,
]{revtex4-1}

\usepackage{physics,graphicx,xcolor}
\usepackage{colortbl}

\usepackage{dsfont}

\newcommand{\cC}{\mathcal{C}}
\newcommand{\cP}{\mathcal{P}}

\newcommand{\1}{\mathds{1}}

\begin{document}


\title{Wishart and random density matrices: Analytical results for the mean-square Hilbert-Schmidt distance}

\author{Santosh Kumar}
 \email{skumar.physics@gmail.com}
\affiliation{%
 Department of Physics, Shiv Nadar University
Gautam Buddha Nadar, Uttar Pradesh 201314, India
}%


\begin{abstract}
\textcolor{black}{
Hilbert-Schmidt distance is one of the prominent distance measures in quantum information theory which finds applications in diverse problems, such as construction of entanglement witnesses, quantum algorithms in machine learning, and quantum state tomography. In this work, we calculate exact and compact results for the mean square Hilbert-Schmidt distance between a random density matrix and a fixed density matrix, and also between two random density matrices. In the course of derivation, we also obtain corresponding exact results for the distance between a Wishart matrix and a fixed Hermitian matrix, and two Wishart matrices. We verify all our analytical results using Monte Carlo simulations. Finally, we apply our results to investigate the Hilbert-Schmidt distance between reduced density matrices generated using coupled kicked tops.}

\end{abstract}

\pacs{Valid PACS appear here}
\maketitle


\section{\label{SecIntro}Introduction}

The statistical investigation of random density matrices is a very active area of research~\cite{BZ2017,ZS2001,ZS2003,SZ2003,SZ2004,ZPNC2011,CN2016,K2018,Lubkin1978,LP1988,Page1993,B1996,H1998,S1999,BS2001,HLW2006,G2007,OSZ2010,NMV2010,NMV2011,V2010,KP2011,VPO2016,MZB2017,M2007,PPZ2016,RSI2016,BSZW2016,Wei2017,W2019,SK2019,KSA2017,K2019,FK2019,RSI2016}. It not only touches upon some of the fundamental issues in quantum mechanics, but is also crucial to various applications in quantum information processing devices, such as quantum computers, teleporters, cloners, etc~\cite{NC2000,BZ2017,Carolan2015,W2017,MRT2004,MHS2012,BB1984,E1991,R2005,PEWFB2005,LB2000,C2018,BAS2018,Qi2017,Arute2019,TBCL2019}. \textcolor{black}{One of the important aspects in this context concerns with various distance measures between quantum states~\cite{HLW2006,B1996,BZ2017,NC2000,W2017,M2007,JTMA2008,AL2015,PPZ2016,RSI2016,BSZW2016,MZB2017,CCC2019}.} A very important example of practical applicability of these distance measures is in quantifying the accuracy of a signal transmission in quantum communication, wherein one measures the distance between the transmitted and received states~\cite{TBCL2019}. Some examples of widely used distance measures are the trace distance ($d_{\tr}$), Hilbert-Schmidt distance ($d_\mathrm{HS}$), Bures distance ($d_\mathrm{B}$), and Hellinger distance ($d_\mathrm{H}$). For given two density matrices $\rho_1,\rho_2$, these are defined respectively as~\cite{H1992,J1994,BZ2017,DLH2011,W2017,NC2000,U1976,B1969,BTL2019,GLN2005,LZ2004}
\begin{align*}
&d_{\tr}=\tr|\rho_1-\rho_2|,\\
&d_\mathrm{HS}=\sqrt{\tr|\rho_1-\rho_2|^2},\\
&d_\mathrm{B}=\sqrt{2-2\tr(\sqrt{\rho_1}\rho_2\sqrt{\rho_1})^{1/2}},\\
&d_\mathrm{H}=\sqrt{2-2\tr(\sqrt{\rho_1}\sqrt{\rho_2})}.
\end{align*}
Here, `tr' represents trace and $|A|$ for a given matrix or operator $A$ is defined as the positive square root of $A^\dag A$, i.e. $|A|=\sqrt{A^\dag A}$. Often, some additional numerical factors are introduced in the above definitions to fix desired normalizations. It may be noted that for density matrices we have $|\rho_1-\rho_2|^2=(\rho_1-\rho_2)^2$, since they are Hermitian. \textcolor{black}{Trace distance possesses the contractivity property, however it is non-Riemannian. Hilbert-Schmidt distance is Riemannian, but not contractive (or, equivalently, monotone) in general. Bures and Hellinger distances are both Riemannian and monotone. These and other properties exhibited by these distance measures lead to corresponding interesting physical consequences, and accordingly their suitability for various applications in quantum information theory is decided~\cite{H1992,J1994,LZ2004,U1976,B1969,BTL2019,GLN2005,BZ2017,NC2000,DLH2011,RSI2016,W2017,CFS2019,DMMW2000,TBCL2019}.}   

\textcolor{black}{Hilbert-Schmidt distance has been one of the prominent and natural choices for quantifying the separation between given two quantum states~\cite{BTL2019,TBCL2019,DMMW2000,CFS2019,LKB2003,BNT2002,BDHK2005,BK2008,PSW2019,WPSW2020,LTOCC2019,ACSZC2019,KLPCSC2019,CPCC2019,Scott2008,ZE2011,STM2013,KKF2020,AG2011,RSI2016,Gao2016,HHW2018,LZ2019}. It provides a direct interpretation as an information distance between quantum states~\cite{LKB2003}. It plays a crucial role in connection to entanglement witness operators~\cite{BNT2002,BDHK2005,BK2008}, being equal to the maximal violation of the associated inequality. A recent example in this context is its implementation in the Gilbert algorithm~\cite{G1966} to construct entanglement witnesses for unextendible product basis bound entangled states~\cite{PSW2019,WPSW2020}.  Moreover, Hilbert-Schmidt distance has been utilized as a cost function in variational hybrid quantum-classical algorithms in machine learning and other applications~\cite{LTOCC2019,ACSZC2019,KLPCSC2019,CPCC2019,TBCL2019}. It has been regularly employed as an estimator in the precision quantum-state tomography~\cite{Scott2008,ZE2011,STM2013,KKF2020}. It also finds applications in the calculation of nonclassical correlations between quantum states other than entanglement, such as quantum discord~\cite{AG2011,RSI2016,Gao2016,HHW2018,LZ2019}. As far as distinguishability criterion is concerned, Hilbert-Schmidt distance does have its limitations since it does not possess contractivity property in general~\cite{NC2000,BZ2017,W2017,O2000,WS2009}. However, archetype quantum systems such as qubits constitute useful exceptions where contractivity is retained and the Hilbert-Schmidt distance equals the trace-distance up to a constant factor~\cite{DLH2011}. Finally, a strong bound between trace distance and Hilbert-Schmidt distance is now known due to Ref.~\cite{CCC2019}. }

Several researchers have worked on the aforementioned distance measures\textcolor{black}{, including Hilbert-Schmidt,} in the context of random density matrices. For instance, in Refs.~\cite{PPZ2016,MZB2017} the authors have derived, {\it inter alia}, averages of the above distances between two Hilbert-Schmidt distributed random density matrices in large matrix-dimension limit using free probability techniques~\textcolor{black}{\cite{MR2017,RE2008}}. The average distance of random states from maximally entangled and coherent states has been calculated in Ref.~\cite{BSZW2016}. These results involving the random density matrices serve as references with which one can compare the distances between quantum states of interest~\cite{PPZ2016,MZB2017,BSZW2016}. \textcolor{black}{This kind of statistical approach is adequate in view of the typicality exhibited by various quantities in quantum information theory. An example is the typicality of quantum entanglement exhibited by random bipartite pure states sampled using the unitarily invariant Haar measure~\cite{Page1993,HLW2006,ODP2007,DLMS2014,ZSP2017}. The underlying phenomenon is that of concentration of measure and such typical behavior conform to the \emph{equal a priori} postulate of the statistical physics~\cite{Tasaki1998,GLTZ2006,PSW2006,IB2017}.}

\textcolor{black}{Exact and finite Hilbert-space dimension results hold a special place in quantum information theory and are especially suited for dealing with real world experiments~\cite{Carolan2015,MRT2004,MHS2012,Qi2017,TBCL2019,PEWFB2005,C2018,Arute2019}. A prominent example is the seminal result of Page for the average von Neumann entropy associated with the subsystems of a composite bipartite system~\cite{Page1993}. This result has found application in diverse problems, including many-body localization in spin systems~\cite{YCHM2015}, entanglement in neural network states~\cite{DLS2017}, and information in black hole radiation~\cite{Page1993a}.}

In this work, we derive exact and compact results for the mean square Hilbert-Schmidt distance, i.e., the average of squared Hilbert-Schmidt distance,
\begin{align*}
D^2:=\mathbb{E}\big[d_\mathrm{HS}^2]=\mathbb{E}\big[\tr(\rho_1-\rho_2)^2\big],
\end{align*}
where the average $\mathbb{E}[\,\cdot\,]$ is with respect to the probability measure governing the random density matrices. To this end, we use the relationship between the Wishart random matrix ensemble and the corresponding fixed trace variant. The latter serves as a model for describing random density matrices. 
To begin with, in Sec.~\ref{SecWis}, we derive exact results for the average of squared Hilbert-Schmidt distance between a random matrix taken from the Wishart ensemble and a fixed Hermitian matrix, and also between two Wishart random matrices. These results are then used in Sec.~\ref{SecDM} to compute exact results for the mean square Hilbert-Schmidt distance between a random density matrix taken from the set of density matrices equipped with the Hilbert-Schmidt measure~\cite{ZS2001,SZ2004} and a fixed density matrix, and also between two random density matrices. We verify all our analytical results using Monte Carlo simulations.  In Sec.~\ref{SecCKT}, we evaluate the mean square Hilbert-Schmidt distance using random density matrices generated via coupled kicked top systems and compare with our analytical results. Finally, we conclude with a brief summary and outlook in Sec.~\ref{SecConc}.


\section{\label{SecWis} Mean square Hilbert-Schmidt distance for Wishart matrices}

The probability density function associated with the Wishart (or Wishart-Laguerre) random matrices is given by~\cite{F2010,W1928,GN1999,A2003}
\begin{equation}
\label{PW}
P(W)=C (\det W)^\alpha e^{-\frac{\beta}{2}\tr W},
\end{equation}
where `$\det$' represents determinant and, as mentioned earlier, `$\tr$' is the trace. The parameter $\alpha$ is decided by the Dyson index $\beta$, the dimension $n$ and the number of degrees of freedom $m$,
\begin{equation}
\alpha=\frac{\beta}{2}(m-n+1)-1.
\end{equation} 
For $\beta=1$ the random matrix $W$ is real positive-definite and for $\beta=2$ it is complex-Hermitian positive-definite.
The inverse of the normalization constant $C$ (partition function) is given by
\begin{equation}
\label{Cinv}
C^{-1}=\left(\frac{2}{\beta}\right)^{\beta n m/2}\pi^{\beta n(n-1)/4} \prod_{i=1}^n \Gamma\left(\frac{\beta}{2}(m-i+1)\right).
\end{equation}
The Wishart matrix $W$ of Eq.~\eqref{PW} can be constructed as
\begin{equation}
W=GG^\dag,
\end{equation}
where $G$ is an $n\times m$-dimensional real (for $\beta=1$) or complex (for $\beta=2$) Ginibre-Gaussian random matrix from the distribution
\begin{equation}
\label{PG}
P_G(G)=\left(\frac{\beta}{2\pi}\right)^{\beta n m/2} e^{-\frac{\beta}{2}\tr (GG^\dag)}.
\end{equation}
Here, `$\dag$' represents transpose and conjugate-transpose for $\beta=1$ and 2, respectively. 

In the following subsections, we derive the desired averages for squared Hilbert-Schmidt distance.

\subsection{Wishart matrix and a fixed matrix}

Let $W$ be an $n$-dimensional Wishart random matrix from the distribution given in Eq.~\eqref{PW}. Also, consider $X$ to be a fixed $n$-dimensional real-symmetric (for $\beta=1$) or complex-Hermitian (for $\beta=2$) matrix. We are interested in calculating the average of the squared Hilbert-Schmidt distance between $W$ and $X$. It can be calculated as
\begin{align}
\nonumber
&D^2_{W,X}=\int d[W]\, P(W)\tr(W-X)^2 \\
\nonumber
&=\int d[W]\,P(W) \tr W^2 +\int d[W]P(W) \tr X^2\\
&~  -2\int d[W]\,P(W)\tr(W X).
\end{align}
Here, $d[W]$ represents the differential of all the independent components in $W$, i.e., \textcolor{black}{$d[W]=\prod_{j\le k}dW_{jk}$ for $\beta=1$ and $d[W]=\prod_i W_{ii} \prod_{j<k} d\text{Re}(W_{jk})d\text{Im}(W_{jk})$ for $\beta=2$. Here, in the $\beta=2$ case, `Re' and `Im' represent the real and imaginary parts of the off-diagonal elements of $W$, which happen to be complex variables.} The average of $\tr W^2$ is well known in the existing literature for both real and complex cases; see for example Refs.~\cite{M1986,NG2011}. Alternatively, it can be also obtained by calculating the corresponding average using the eigenvalues of $W$ with the aid of Selberg integrals~\cite{M2004,F2010}. We obtain 
\begin{align}
\label{EtrW2}
&\int d[W]\,P(W) \tr W^2=n m(n+m+2/\beta-1).
\end{align}
\textcolor{black}{We note that $ \tr W^2$ is the second spectral moment of the random matrix $W$, and therefore the above integral gives its mean value.}
It is also known that~\cite{M1986,NG2011}
\begin{equation}
\label{EtrWX}
\int d[W] P(W)\tr(WX)=m\,\tr X.
\end{equation}
\textcolor{black}{The above can be viewed as the mean scalar (inner) product between the random matrix $W$ and the fixed matrix $X$.}
Now, we have
\begin{align}
\label{D2WX}
\nonumber
&D^2_{W,X}=n m(n+m+2/\beta-1)+\tr X^2-2m\,\tr X\\
&=n m(n+m+2/\beta-1)+\sum_{i=1}^n \chi_i(\chi_i-2m),
\end{align}
where $\chi_i$ are the eigenvalues of $X$. \textcolor{black}{The above result holds even if we consider $X\to zX$ with $z$ being some complex scalar. It should be noted, however, that in this case $zX$ is not a real-symmetric or complex-Hermitian matrix in general.}

\begin{figure}[h!]
\centering
\includegraphics[width=0.9\linewidth]{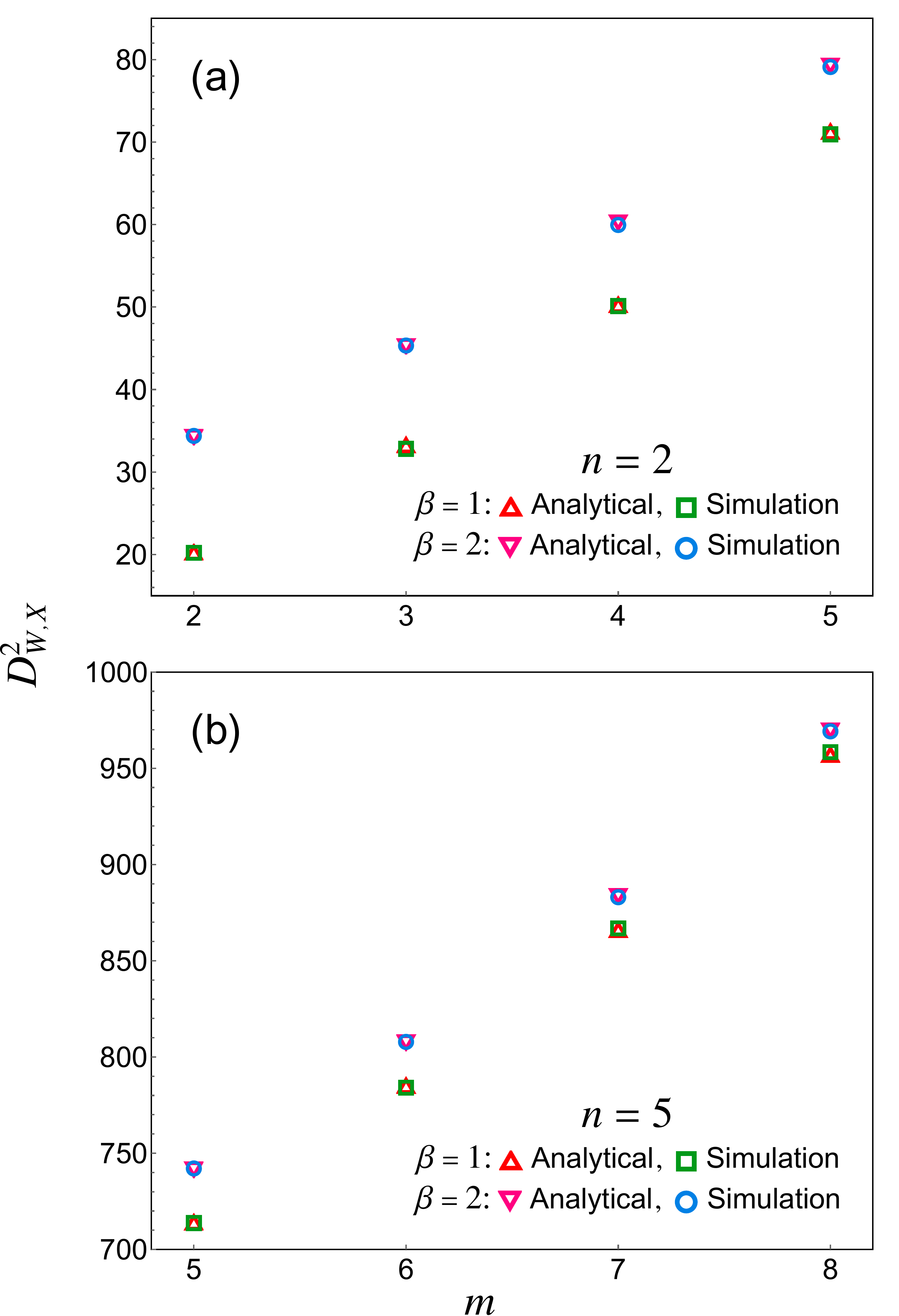}
\caption{Mean-square Hilbert-Schmidt distance between a Wishart matrix and a fixed matrix for (a) $n=2$ and (b) $n=5$. The $m$ (number of degrees of freedom) value for the Wishart matrix varies from $n$ to $n+3$ in both cases.}
\label{fig1}
\end{figure}

We compare the above analytical result with averages obtained using Monte Carlo simulation involving $10^5$ Wishart matrices for both $\beta=1$ and $2$ cases. We consider $n=2,5$, and $m$ varying from $n$ to $n+3$. The fixed matrix $X$ chosen in the $n=2$ and 5 cases are
\begin{align*}
&~~\begin{pmatrix}
2 & 1\\ 1 & -1/2
\end{pmatrix}, & \beta=1,\\
&\begin{pmatrix}
2 & 1+3i\\ 1-3i & -1/2
\end{pmatrix}, & \beta=2,
\end{align*}
and
\begin{align*}
&~~~~~~~~~~\begin{pmatrix}
3 & 1 & 4 & 6 & 8\\ 1 & -5 & 4 & 7 & -1 \\ 4 & 4 & 2 & 1 & 3 \\ 6 & 7 & 1 & 9 & 0 \\ 8 & -1 & 3 & 0 & -2
\end{pmatrix}, & \beta=1,\\
&\begin{pmatrix}
3 & 1 + i & 4 - i/2 & 6 +  \sqrt{3}\,i & 8 - i\\ 1 - i & -5 & 4 + 3 i & 7 & -1\\ 4 + i/2 & 4 - 3 i & 2 & 2 - 3 i & 3 \\ 6 -  \sqrt{3}\,i & 7 & 2 + 3 i & 9 & i/5 \\ 8 + i & -1 & 3 & -i/5 & -2
\end{pmatrix}, & \beta=2,
\end{align*}
respectively. The comparison is shown with the aid of various symbols in Fig.~\ref{fig1} and we observe that the analytical and simulation based results agree very well.

\subsection{Two Wishart matrices}

We now consider two $n$-dimensional Wishart-Laguerre matrices $W_1$ and $W_2$ but with different choices for the number of degrees of freedom in general, say $m_1$ and $m_2$, i.e., we consider the respective probability densities as $P_j(W_j)=C_j (\det W_j)^{\frac{\beta}{2}(m_j-n+1)-1}e^{-\frac{\beta}{2}\tr W_j}$; $j=1,2$. The average of the squared Hilbert-Schmidt distance between $W_1$ and $W_2$ then follows as
\begin{align}
D^2_{W_1,W_2}=\int d[W_1]\int d[W_2]P_1(W_1)P_2(W_2) \tr(W_1-W_2)^2.
\end{align}
We can evaluate the $W_2$ integral first by keeping $W_1$ fixed and using Eq.~\eqref{D2WX}. This gives us
\begin{align}
\nonumber
D^2_{W_1,W_2}=\int d[W_1]P_1(W_1) \Big[n m_2(n+m_2+2/\beta-1)\\
+\tr W_1^2-2m_2\,\tr W_1\Big] .
\end{align}
Now, the integral over the first term in the above expression is trivial, the second term can be integrated using Eq.~\eqref{EtrW2}, and the third term can be integrated using Eq.~\eqref{EtrWX} with $X=\mathds{1}_n$. We obtain the desired expression as
\begin{align}
\label{D2WW}
\nonumber
&D^2_{W_1,W_2}=n m_1(n+m_1+2/\beta-1)\\
\nonumber
&~~+n m_2(n+m_2+2/\beta-1)-2n m_1m_2\\
&=n\left[(m_1+m_2)(n+2/\beta-1)+(m_1-m_2)^2\right].
\end{align}

\begin{figure}[h!]
\centering
\includegraphics[width=0.9\linewidth]{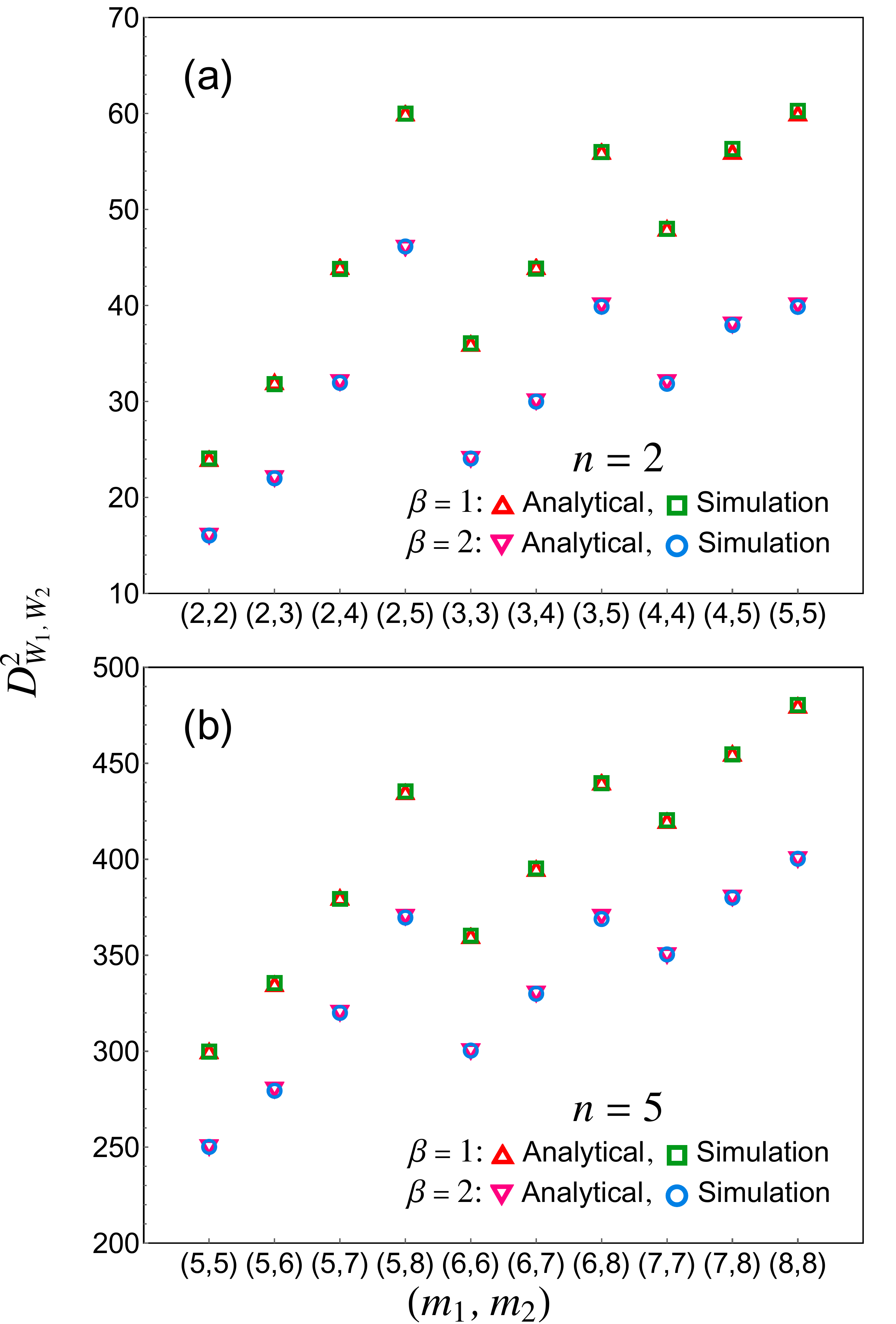}
\caption{Mean-square Hilbert-Schmidt distance between two independent Wishart matrices of dimension (a) $n=2$ and (b) $n=5$. In both cases, various combinations of the numbers of degrees of freedom $m_1$ and $m_2$ have been considered.}
\label{fig2}
\end{figure}

The above result is verified using Monte Carlo simulations involving $10^5$ pairs of Wishart matrices. In Fig.~\ref{fig2}, we show the comparison for $n=2$ and $5$ with various combinations of $m_1$ and $m_2$ as indicated. We can see a very good agreement in all cases.

\section{\label{SecDM} Mean square Hilbert-Schmidt distance for random density matrices}

\textcolor{black}{We now focus on random density matrices taken from the set equipped with the Hilbert-Schmidt measure.} The corresponding probability density function is given by~\cite{ZS2001},
\begin{equation}
\label{Prho}
\cP(\rho)=\cC (\det\rho)^{\alpha}\delta(\tr \rho-1).
\end{equation}
As shown in the Appendix, the normalization factor $\cC$ in this case is related to the one in Eq.~\eqref{Cinv} as
\begin{equation}
\label{CcC}
\cC=\left(\frac{2}{\beta}\right)^{\beta n m/2} \Gamma(\beta n m/2) C.
\end{equation}
The $\beta=1$ case here can be associated with real random states, while $\beta=2$ corresponds to the usual scenario of complex states~\cite{ZS2001}.

The above described measure over random density matrices arises in the following way. Consider a random pure state $\ket{\psi}$ belonging to the Hilbert space $\mathcal{H}_n\otimes\mathcal{H}_m$ which is associated with a composite bipartite system of size $nm$ with $n\le m$. \textcolor{black}{This random pure state may be represented as $\ket{\psi}=U\ket{\psi_0}$, where $U$ is a global random unitary matrix distributed according to the Haar measure, and $\ket{\psi_0}$ is an arbitrary state in $\mathcal{H}_n\otimes\mathcal{H}_m$.} Upon partial tracing over the $m$-dimensional \emph{environment} part, one obtains the reduced density matrix of dimension $n$,
\begin{align}
\label{rdm}
\rho=\frac{\tr_m(\ket{\psi}\bra{\psi})}{\bra{\psi}\ket{\psi}}.
\end{align}
This reduced density matrix is then distributed as described by the probability density in Eq.~\eqref{Prho}~\cite{ZS2001}. The $n=m$ case is identified as the standard Hilbert-Schmidt measure and is also induced by the Hilbert-Schmidt distance metric~\cite{ZS2001}. The construction appearing in Eq.~\eqref{rdm} maps to the random matrix model~\cite{ZS2001,SZ2004,OSZ2010,ZPNC2011} 
\begin{equation}
\label{FT}
\rho=W/\tr W=GG^\dag/\tr(GG^\dag),
\end{equation}
where $W$ and $G$ are matrices as in Eqs.~\eqref{PW} and~\eqref{PG}. \textcolor{black}{Evidently, this results in the random matrix $\rho$ having a fixed trace 1 and therefore, in the random matrix theory terminology, it is said to belong to the fixed trace Wishart-Laguerre ensemble~\cite{ZS2001,SZ2004,F2010,OSZ2010,ZPNC2011}. We exploit the above relationship between the random density matrix $\rho$ and the Wishart matrix $W$ to obtain the mean square Hilbert-Schmidt distances for the former with the help of results derived in the preceding section.}

\subsection{A random density matrix and a fixed density matrix}

Let $\rho$ be a random density matrix from the distribution given in Eq.~\eqref{Prho} and $\sigma$ be a fixed density matrix. We need to calculate average of the squared Hilbert-Schmidt distance between $\rho$ and $\sigma$,
\begin{align}
&D^2_{\rho,\sigma}=\int d[\rho]\, \cP(\rho)\tr(\rho-\sigma)^2,
\end{align}
\textcolor{black}{where $d[\rho]$ is defined similar to $d[W]$.} We introduce an auxiliary variable $t$ inside the delta function to replace 1 in the expression of the density $\cP(\rho)$. It will be set equal to 1 towards the end of the calculation. We have
\begin{align}
D^2_{\rho,\sigma}(t)=\cC\int d[\rho]\,(\det\rho)^{\alpha}\delta(\tr \rho-t) \tr(\rho-\sigma)^2 .
\end{align}
Taking Laplace transform ($t\to s$), we get
\begin{align}
\widetilde{D^2_{\rho,\sigma}}(s)=\cC\int d[\rho]\,(\det\rho)^{\alpha}e^{-s\,\tr \rho} \tr(\rho-\sigma)^2.
\end{align}
We now introduce $\rho=(\frac{\beta}{2s})W$ with $s>0$, so that $d[\rho]=(\frac{\beta}{2s})^{n[\beta(n-1)/2+1]}d[W]$. After some simplification we obtain
\begin{align}
\nonumber
&\widetilde{D^2_{\rho,\sigma}}(s)
=\cC\left(\frac{\beta}{2s}\right)^{\beta nm/2+2}\int d[W]\,(\det W)^{\alpha}e^{-\,\frac{\beta}{2}\tr W} \\
\nonumber
&~~~~~~~~~~~~~~~\times \tr\left(W-\frac{2s}{\beta}\sigma\right)^2\\
\nonumber
&=\frac{\cC}{C}\left(\frac{\beta}{2s}\right)^{\beta nm/2+2}\int d[W]P(W) \tr\left(W-\frac{2s}{\beta}\sigma\right)^2\\
\nonumber
&=\frac{\cC}{C}\left(\frac{\beta}{2s}\right)^{\beta nm/2+2}\big[n m(n+m+2/\beta-1)\\
&~~~~~~~~~~~ +(4s^2/\beta^2)\tr\sigma^2-(4m/\beta)s\,\tr \sigma\big],
\end{align}
where we employed Eqs.~\eqref{PW} and~\eqref{D2WX}.
Now, $\sigma$ being a density matrix, we have $\tr\,\sigma=1$. Taking the inverse Laplace transform ($s\to t$) then yields
\begin{align}
\nonumber
&D^2_{\rho,\sigma}(t)=\frac{\cC}{C}\left(\frac{\beta}{2}\right)^{\beta nm/2+2}\bigg[\frac{4t^{\beta nm/2-1}}{\beta^2\Gamma(\beta nm/2)}\tr \sigma^2\\
\nonumber
& - \frac{4m\,t^{\beta nm/2}}{\beta\Gamma(\beta nm/2+1)}
+n m(n+m+2/\beta-1)\frac{t^{\beta nm/2+1}}{\Gamma(\beta nm/2+2)}\bigg].
\end{align}
Finally, setting $t=1$ and substituting the ratio $\cC/C$ from Eq.~\eqref{CcC}, we obtain the desired result:
\begin{align}
\label{Drhosig}
D^2_{\rho,\sigma}=\tr \sigma^2+\frac{\beta(n+m+2/\beta-1)}{\beta n m+2}-\frac{2}{n}.
\end{align}
The above derivation, equivalently, may be carried out by observing that $\mathcal{P}(\rho)\propto\int d[G]\delta(\rho-GG^\dag)\delta(\tr GG^\dag-1)P_G(G)$ and mapping the $\rho$-integral to $G$-integral.
It should be noted that the second term in Eq.~\eqref{Drhosig} corresponds to the average of $\tr \rho^2 $, i.e., it is the average purity for a random density matrix, viz.
\begin{equation}
\label{avpur}
\int d[\rho]\, \cP(\rho)\tr \rho^2=\frac{\beta(n+m+2/\beta-1)}{\beta n m+2}.
\end{equation}
Of special interest is the case when $\sigma$ is a pure state or a maximally mixed state. For these, we have $\tr \sigma^2=1$ and $1/n$, respectively and the corresponding average distances can be readily obtained from Eq.~\eqref{Drhosig}. Moreover, for $m=n\gg1$, we obtain
\begin{equation}
D^2_{\rho,\sigma}= \tr \sigma^2+\mathcal{O}\left(\frac{1}{n^2}\right),
\end{equation}
which, to the leading order, is just the purity of the state $\sigma$.
In the same limit, the leading contribution for pure and maximally-mixed states are therefore $D^2_{\rho,\sigma}=1$ and $D^2_{\rho,\sigma}=1/n$. The latter goes to 0 as $n\to\infty$, as was shown in Ref.~\cite{HLW2006}.

\begin{figure}[h!]
\centering
\includegraphics[width=0.9\linewidth]{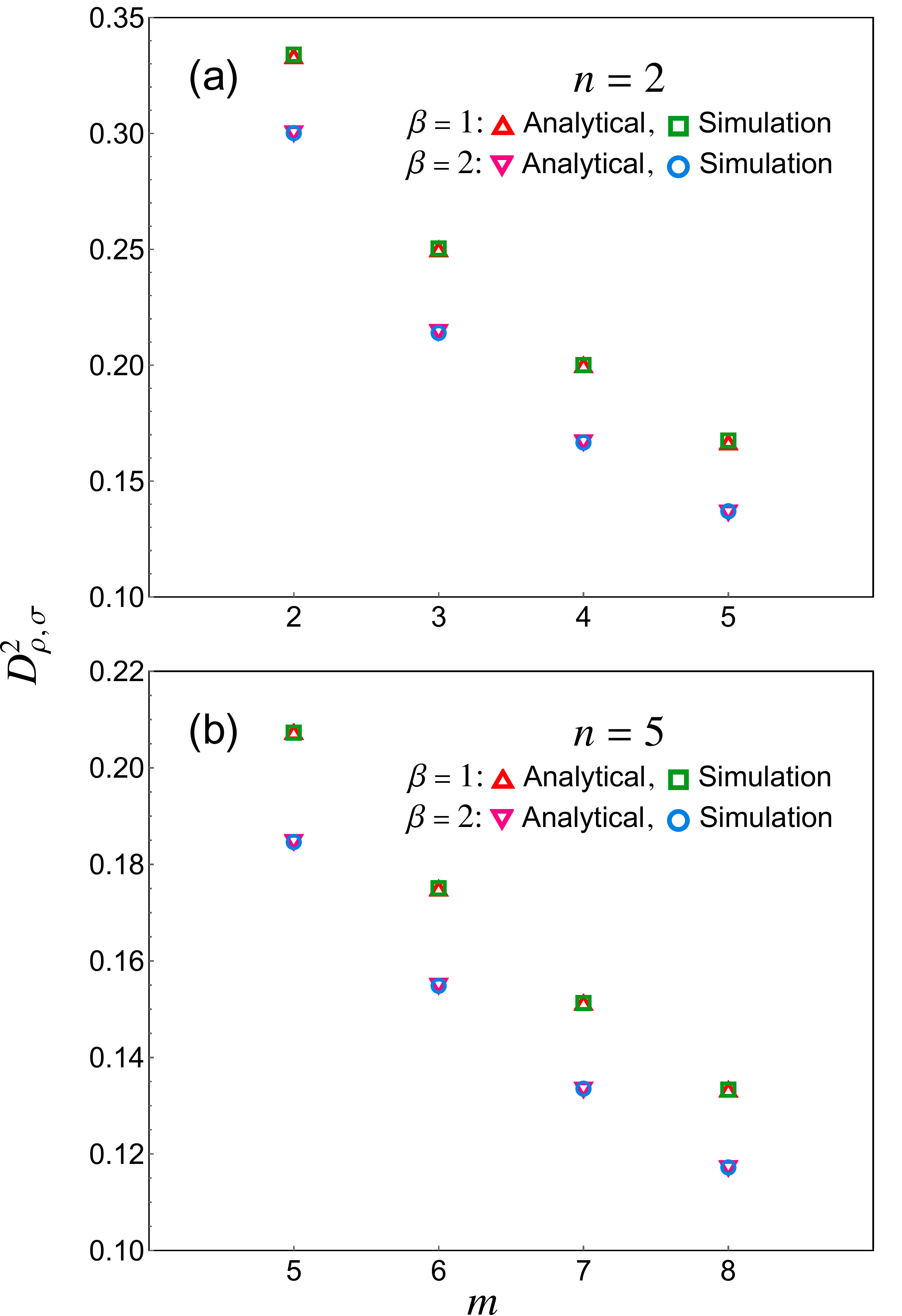}
\caption{Mean-square Hilbert-Schmidt distance between a random density matrix and a fixed density matrix for (a) $n=2$ and (b) $n=5$. The $m$ (Hilbert space dimension of the environment) value varies from $n$ to $n+3$ in both cases.}
\label{fig3}
\end{figure}

We verify Eq.~\eqref{Drhosig} by numerically simulating $10^5$ random density matrices using the random matrix model, Eq.~\eqref{FT}, and calculating the mean distance square with the fixed matrix $\sigma$ set as the maximally mixed state $n^{-1}\1_n$. The results are depicted in Fig.~\ref{fig3} for $n=2,5$, and $m$ varying from $n$ to $n+3$. We find an impressive agreement between the analytical and simulation based results. 

\subsection{Two random density matrices}

Let $\rho_1$ and $\rho_2$ be random density matrices from the probability density given in Eq.~\eqref{Prho}, but unequal $m$ in general, say $m_1$ and $m_2$. We therefore need to calculate
\begin{align}
&D^2_{\rho_1,\rho_2}=\int d[\rho_1]\int d[\rho_2]\,\cP_1(\rho_1)\,\cP_2(\rho_2) \tr(\rho_1-\rho_2)^2 ,
\end{align}
where $\cP_j(\rho_j)=\cC_j (\det\rho_j)^{\frac{\beta}{2}(m_j-n+1)-1}\delta(\tr \rho_j-1); j=1,2$.
We can calculate the $\rho_2$ integral first by treating $\rho_1$ fixed, and thus use Eq.~\eqref{Drhosig}. We obtain
\begin{align*}
D^2_{\rho_1,\rho_2}=\int d[\rho_1] \cP_1(\rho_1)\Big[\tr\rho_1^2+\frac{\beta(n+m_2+2/\beta-1)}{\beta n m_2+2}-\frac{2}{n}\Big].
\end{align*}
The first term can be integrated using Eq.~\eqref{avpur}, while the integral over the other two terms is trivial. We have
\begin{align}
\label{Drhosgm}
\nonumber
D^2_{\rho_1,\rho_2}&=\frac{\beta(n+m_1+2/\beta-1)}{\beta n m_1+2}\\
&+\frac{\beta(n+m_2+2/\beta-1)}{\beta n m_2+2}-\frac{2}{n}.
\end{align}
For $n=m_1=m_2\gg1$, we obtain
\begin{equation}
D^2_{\rho_1,\rho_2}=\frac{2}{n}+\mathcal{O}\left(\frac{1}{n^2}\right),
\end{equation}
as was calculated in Ref.~\cite{PPZ2016}.

\begin{figure}[h!]
\centering
\includegraphics[width=0.9\linewidth]{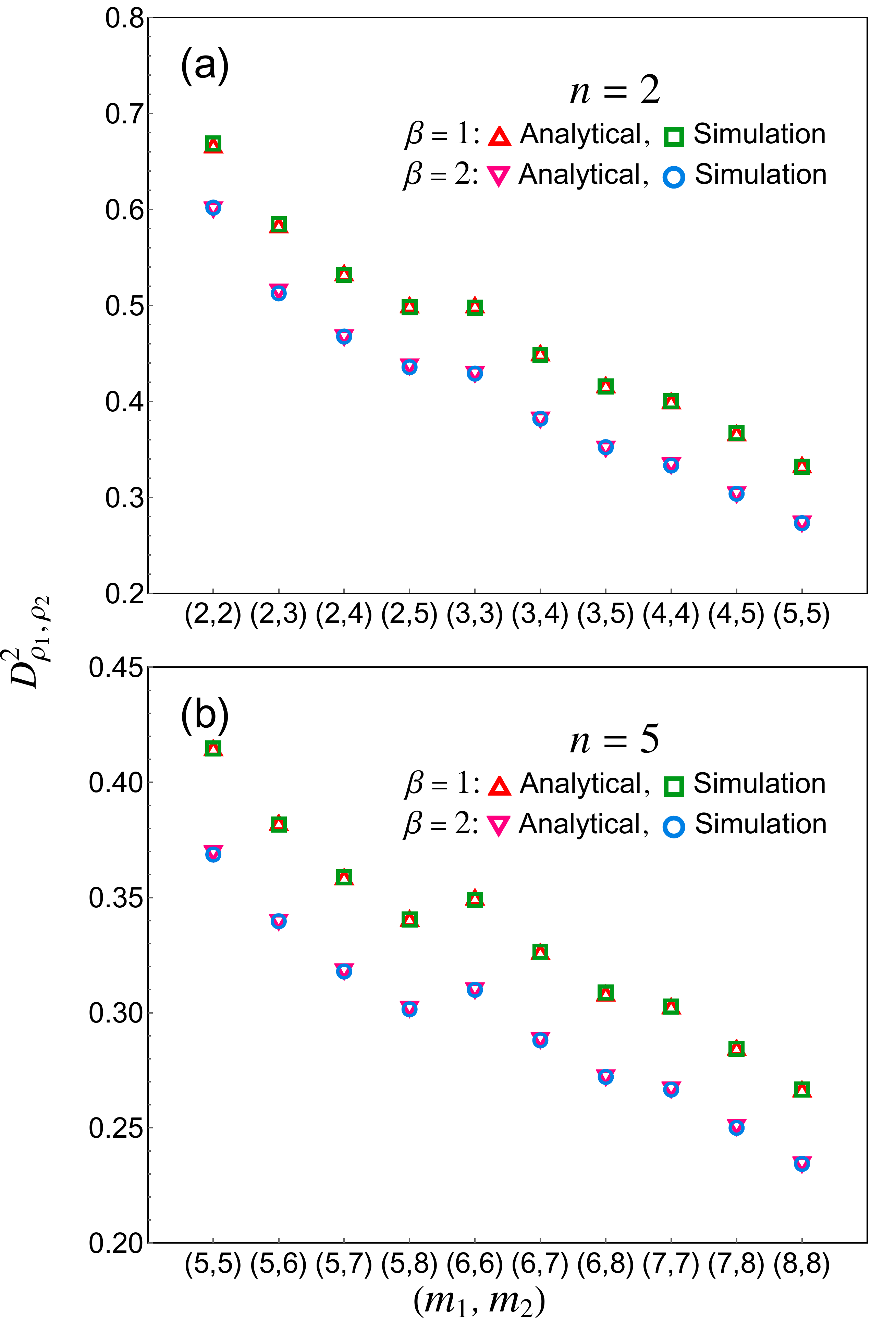}
\caption{Mean-square Hilbert-Schmidt distance between two independent random density matrices with (a) $n=2$ and (b) $n=5$. For both cases, several combinations of  $m_1,m_2$ values have been considered.}
\label{fig4}
\end{figure}

We simulate $10^5$ pairs of random density matrices using the matrix model in Eq.~\eqref{FT} and obtain the average of Hilbert-Schmidt distance square. These Monte Carlo results are contrasted with the above analytical result in Fig.~\ref{fig4}. We have considered $n=2,5$ and several $m_1,m_2$ values and very good agreement can be seen in all cases. 


\begin{figure*}[ht!]
\centering
\includegraphics[width=0.8\linewidth]{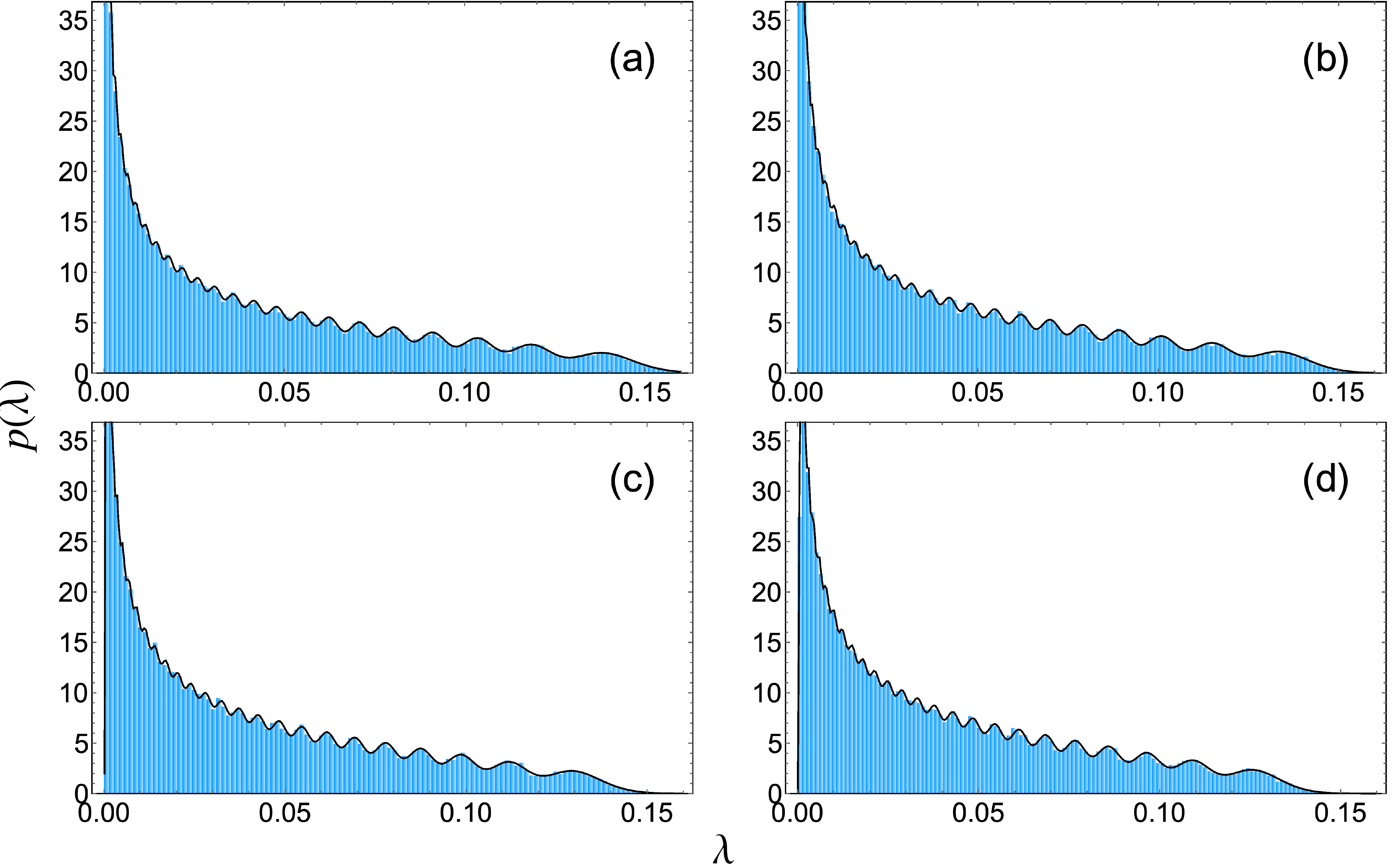}
\caption{Eigenvalue density for Hilbert-Schmidt distributed random density matrix: comparison between analytical result (solid line) and the histogram obtained by diagonalising reduced density matrices obtained from coupled kicked top simulation with $k_1=k_2=7,\epsilon=1$. The density matrix dimension is $n=2j_1+1=25$ and the subplots depict the densities for varying $m=2j_2+1$ values: (a) 25, (b) 27, (c) 29, and (d) 31.}
\label{fig5}
\end{figure*}

\section{\label{SecCKT} Coupled kicked tops}

In this section, we compare the analytical results obtained in the preceding section with the mean square Hilbert-Schmidt distance obtained using random density matrices generated via coupled kicked tops. Coupled quantum kicked tops, {\it inter alia}, have been used extensively to study the bipartite entanglement and effect of chaos~\cite{MS1999,BL2002,FMT2003,BL2004,DK2004,TMD2008,KAT2013,KSA2017}. In Ref.~\cite{PPZ2016}, it has been used to study the spectrum of the difference of two density matrices, the so called Helstrom matrix. In the same spirit, we use here the couple kicked top system to generate random density matrices distributed according to Hilbert-Schmidt measure and then evaluate the corresponding squared Hilbert-Schmidt distance averages. These results are compared with our random matrix theory based analytical results.

The Hamiltonian for the coupled kicked top system is~\cite{MS1999,BL2002}
\begin{equation}
\label{CKTH}
H=H_1\otimes \1_{N_2}+\1_{N_1}\otimes H_2+H_{12}.
\end{equation}
Here, 
\begin{equation}
H_r=\frac{\pi}{2}J_{y_r}+\frac{k_r}{2j_r}J_{z_r}^2\sum_{\nu=-\infty}^{\infty}\delta(t-\nu),~~r=1,2,
\end{equation}
represent the Hamiltonians for the individual tops~\cite{HKS1987,H2010}, and
\begin{equation}
H_{12}=\frac{\epsilon}{\sqrt{j_1j_2}}(J_{z_1}\otimes J_{z_2})\sum_{\nu=-\infty}^{\infty} \delta(t-\nu)
\end{equation}
is the interaction term. The Hamiltonians $H_1$ and $H_2$ correspond to $N_1 ~(=2j_1+1)$-dimensional, and $N_2 ~(=2j_2+1)$-dimensional Hilbert spaces $\mathcal{H}^{(N_1)}$ and $\mathcal{H}^{(N_2)}$, respectively. Also, $\1_{N_1}$ and $\1_{N_2}$ are $N_1$ and $N_2$ dimensional identity operators, respectively. The Hamiltonian for the coupled kicked tops corresponds to an $N_1 N_2$-dimensional Hilbert space $\mathcal{H}^{(N_1N_2)}=\mathcal{H}^{(N_1)}\otimes\mathcal{H}^{(N_2)}$. $J_{x_r}, J_{y_r}, J_{z_r}$ are angular momentum operators for the $r$th top and $j$ is the quantum number corresponding to the operator $J^2$. The stochasticity parameters $k_r$ for the two tops decide the kick strengths and control their chaotic behavior. The parameter $\epsilon$ takes care of the coupling between the two tops.

The unitary time evolution operator (Floquet operator) corresponding to the Hamiltonian in Eq.~\eqref{CKTH} is
\begin{eqnarray}
U=(U_1\otimes U_2)U_{12},
\end{eqnarray}
with
\begin{equation}
U_r=\exp\left(-\frac{\iota\pi}{2}J_{y_r}-\frac{\iota k_r}{2j_r}J_{z_r}^2\right),r=1,2;
\end{equation}
\begin{equation}
U_{12}=\exp\left(-\frac{\iota\epsilon}{\sqrt{j_1j_2}}~J_{z_1}\otimes J_{z_2}\right).
\end{equation}
Here $\iota=\sqrt{-1}$ represents the imaginary unit. The Floquet operator $U$ is used to obtain the state $|\psi(\nu)\rangle$ starting from an initial state $|\psi(0)\rangle$ using the iteration scheme
$|\psi(\nu)\rangle=U|\psi(\nu-1)\rangle$. The initial state is taken as the tensor-product of directed angular momentum states associated with the two tops. After ignoring a certain number of iterations that fall in the transient regime, one considers the reduced density matrices obtained by partial tracing over one of the tops (say, the second one), viz. $\rho(\nu)=\tr_2 (\ket{\psi(\nu)}\bra{\psi(\nu)})$; {\it cf}. Eq.~\eqref{rdm}. In the chaotic regime $(k_r\gtrsim 6)$, with sufficient coupling between the two tops, these reduced density matrices belong to the Hilbert-Schmidt measure as given in Eq.~\eqref{Prho}~\cite{BL2004,KSA2017}. 

\begin{figure}[h!]
\centering
\includegraphics[width=0.9\linewidth]{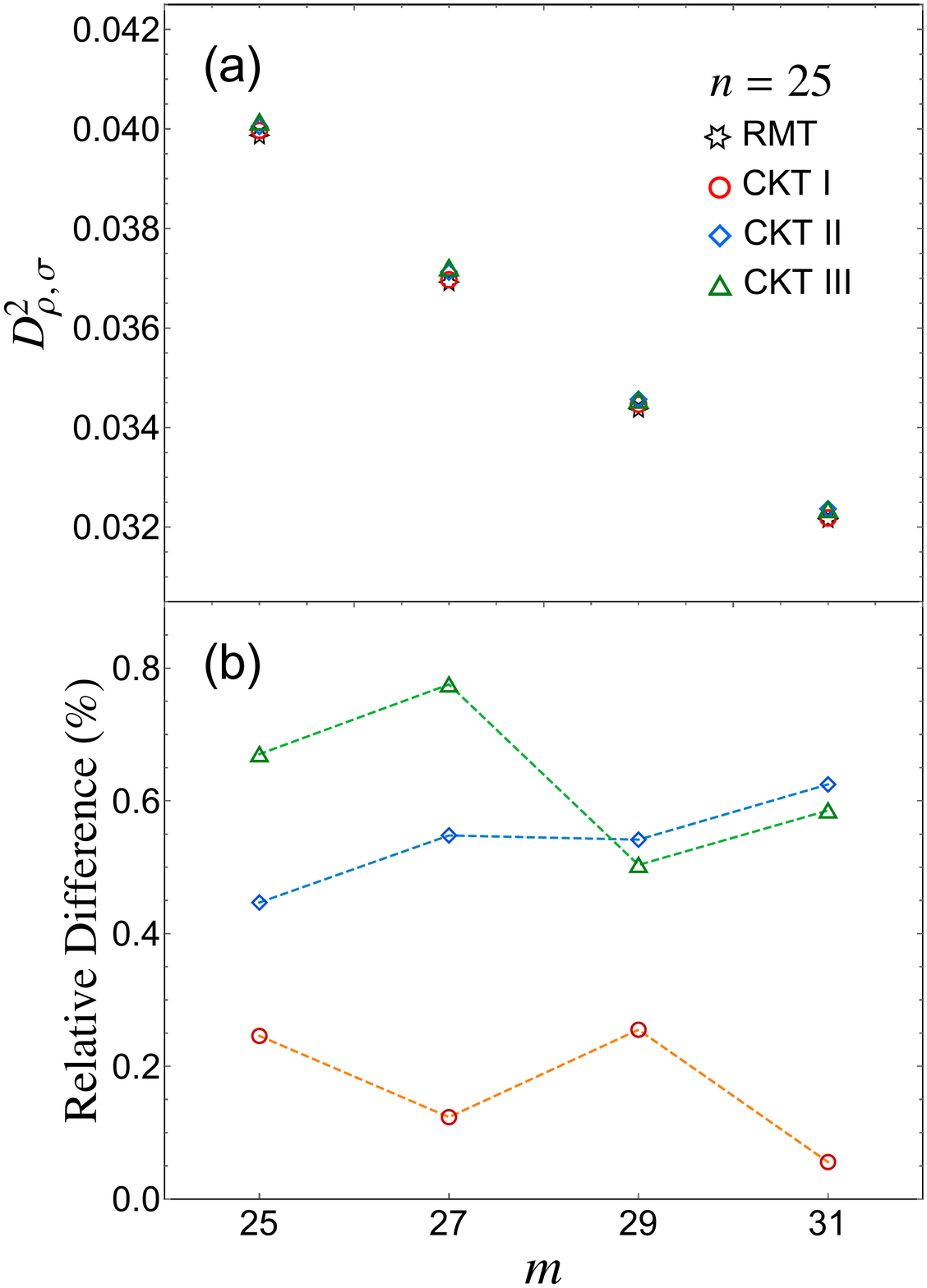}
\caption{\textcolor{black}{Comparison between random matrix theory (RMT) and coupled kicked top simulation results: (a) mean square Hilbert-Schmidt distance between density matrices $\rho$ of dimension $n=25$ generated from coupled kicked top (CKT) simulations and the maximally mixed density matrix $\sigma=n^{-1}\1_n$, along with the RMT predictions; (b) the corresponding percent relative differences, i.e., $100([D_{\rho,\sigma}^2]_\text{CKT}/[D_{\rho,\sigma}^2]_\text{RMT}-1)$\%. The sets of parameters $(k_1,k_2,\epsilon)$ used for the coupled kicked tops are CKT I: $(7,8,1)$, CKT II: $(6,7,0.75)$, CKT III: $(6,9,0.5)$ and $m$ has been varied in each case, as indicated along the horizontal axis.}}
\label{fig6}
\end{figure}

For comparison with our analytical result for distance between a random density matrix and a fixed density matrix, we generate 5000 reduced density matrices using the procedure described above. We consider $j_1=12$ which gives $n=N_1=25$ and vary $j_2$ from 12 to 15 which corresponds to $m=N_2=25,27,29,31$. It should be noted that for each choice of $j_2$, we have to run a separate simulation. The fixed density matrix is chosen as $n^{-1}\1_n$, which represents the maximally mixed state. \textcolor{black}{Before we proceed to calculate the average distance between the quantum states, to demonstrate that the algorithm does produce density matrices distributed according to the Hilbert-Schmidt measure}, we compare the corresponding eigenvalue density with the random matrix prediction for $\beta=2$~\cite{KP2011,KSA2017}, viz.,
\textcolor{black}{
\begin{align}
\nonumber
&p(\mu)=\sum_{i=1}^{n} K_i \,\mu^{i+\alpha-1}(1-\mu)^{-i+n m-\alpha-1}\\
&~~~\times\big[(n-i)\mathcal{F}_{\alpha+1}^{-n,i-nm+\alpha}-n\mathcal{F}_{\alpha+1}^{1-n,i-nm+\alpha}\big].
\end{align}
}
Here $\mu$ represents a generic eigenvalue of $\rho$ and $\mathcal{F}^{a,b}_c:=\,_2F_1(a,b;c;\frac{\mu}{\mu-1})/\Gamma(c)$ with $_2F_1(\cdots)$ being the Gauss hypergeometric function. The coefficient $K_i$ is given by
\textcolor{black}{
\begin{align}
K_i=\frac{(-1)^i\Gamma(m+1)\Gamma(nm)}{n\Gamma(i)\Gamma(n-i+1)\Gamma(i+\alpha+1)\Gamma(nm-\alpha-i)}.
\end{align}
}
As can be seen in Fig.~\ref{fig5}, we find very good agreement between the analytical eigenvalue densities and histograms obtained from simulations. Thus, we use these density matrices for evaluating the Hilbert-Schmidt distance. \textcolor{black}{The results are depicted in Fig.~\ref{fig6} for three sets of $(k_1,k_2,\epsilon)$ parameters along with the random matrix theory based results based on Eq.~\eqref{Drhosig}. We find a very good agreement, with the relative difference remaining below 1\% in each case.}

\begin{figure}[h!]
\centering
\includegraphics[width=0.9\linewidth]{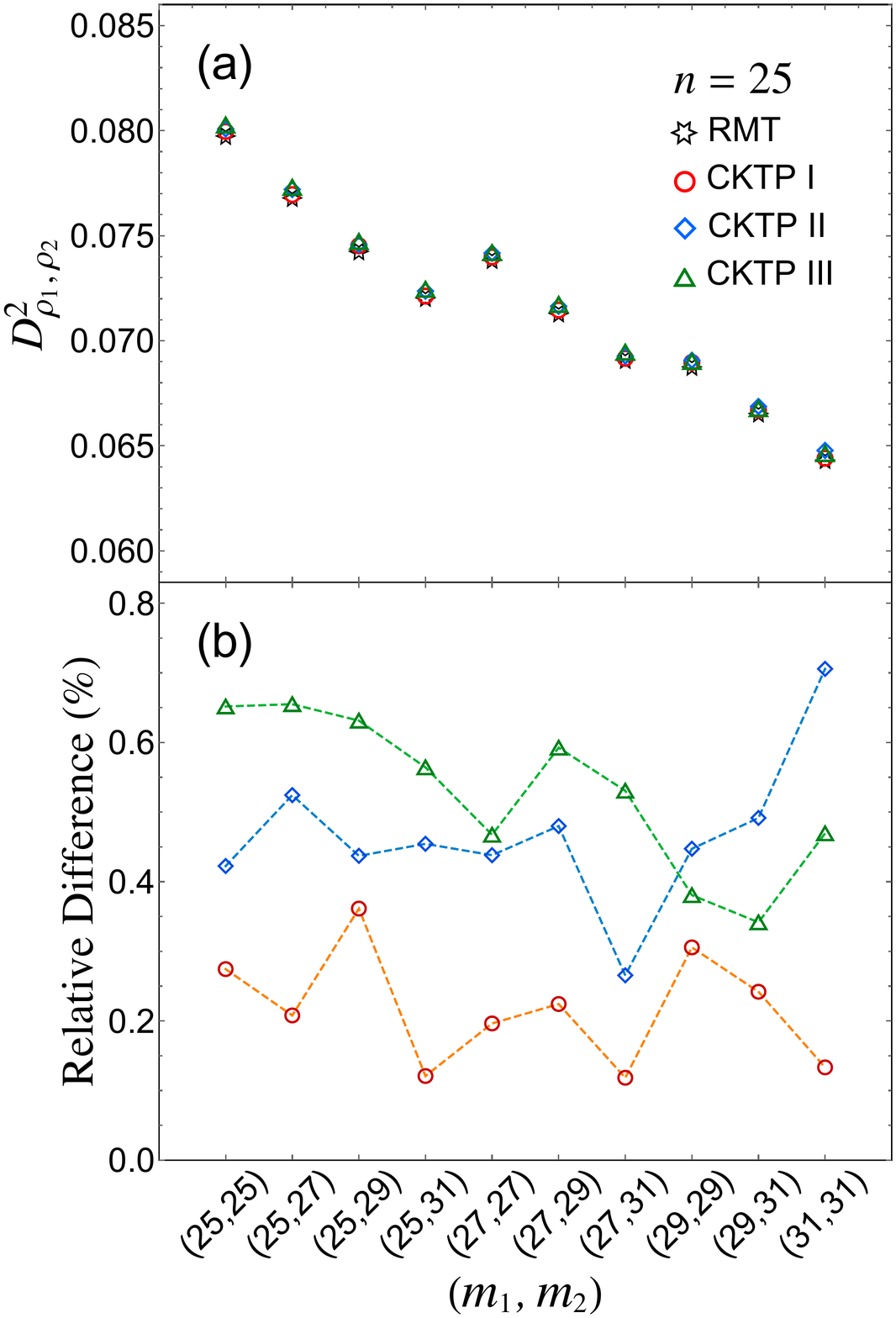}
\caption{\textcolor{black}{Comparison between random matrix theory and coupled kicked top simulation results: (a) mean square Hilbert-Schmidt distance between two random density matrices $\rho_1,\rho_2$ of dimension $n=25$ and various $m_1,m_2$ combinations calculated using coupled kicked top pairs (CKTP), along with RMT predictions; (b) the corresponding percent relative differences. The sets of parameters $(k_1^A,k_2^A,\epsilon^A; k_1^B,k_2^B,\epsilon^B)$ used for the coupled kicked top pairs are CKTP I: $(8,7,0.5;7,8,1)$, CKTP II: $(6,6,0.8;7,8,0.75)$, CKTP III: $(7,7,0.75;8,8,0.75)$.}}
\label{fig7}
\end{figure}

For simulating the distance between two density matrices we consider two independent coupled kicked tops, say $A$ and $B$. This helps us to realize different $m_1=2j_2^{A}+1$ and $m_2=2j_2^{B}+1$ values. Here, $ j_2^{A}$ and $j_2^{B}$ represent the $j_2$ values for the two couple kicked tops, respectively. The $n$ value is decided by the common Hilbert-space dimension $2j_1^{A}+1=2j_1^{B}+1$. We should add that if one does not require to consider different values for $m_1$ and $m_2$, only one coupled kicked top would suffice. In this case, $\rho_1$ and $\rho_2$ can be taken as reduced density matrices separated by a certain number of iterations within a single simulation. \textcolor{black}{In Fig.~\ref{fig7}, we show the comparison between the random matrix analytical and kicked top simulation results for the mean square Hilbert-Schmidt distance for $n=25$ and several combinations of $m_1,m_2$. Three sets of parameters $(k_1^A,k_2^A,\epsilon^A)$ and $(k_1^B,k_2^B,\epsilon^B)$ have been chosen for the coupled tops $A$ and $B$. Here also, we find the agreement to be impressive with the relative difference with the random matrix result, Eq.~\eqref{Drhosgm}, remaining below 1\%.}

\section{\label{SecConc} Summary and outlook}

In this work, we obtained exact and compact expressions for the mean square Hilbert-Schmidt distance between a random density matrix and a fixed density matrix, and also between two random density matrices. This derivation involved first computing the corresponding expressions for Wishart random matrices. \textcolor{black}{These results are compiled in Table 1 for a quick reference.} We also compared our analytical results with the average distances obtained using reduced density matrices simulated via coupled kicked top system with appropriately chosen parameters, and found very good agreement. Our results constitute a useful reference for comparing Hilbert-distance between quantum states. \textcolor{black}{Moreover, due to their simplicity, our analytical expressions are amenable to further analysis, such as examining asymptotic limits.}

\begin{table*}[!ht]
\renewcommand{\arraystretch}{2}
\caption{\textcolor{black}{Summary of results for the mean square Hilbert-Schmidt distance between a pair of matrices. For the Wishart matrices, $n$ is the matrix dimension and $m$ is the number of degrees of freedom. For random density matrices, $n$ is the matrix dimension and $m$ is the auxiliary dimension of the Hilbert-space corresponding to the environment.}}
\centering
{\small \begin{tabular}{ c c }
\arrayrulecolor{black}
\hline\hline
 \textcolor{black}{Matrices} & \textcolor{black}{Mean square Hilbert-Schmidt distance} \\
\hline
\textcolor{black}{A Wishart matrix ($W$) and a fixed Hermitian matrix ($X$)}  & \textcolor{black}{$\displaystyle D^2_{W,X}=n m\left(n+m+\frac{2}{\beta}-1\right)+\tr X^2-2m\,\tr X$} \\
\textcolor{black}{Two Wishart matrices ($W_1,W_2$)} & \textcolor{black}{$\displaystyle D^2_{W_1,W_2}=n\left[(m_1+m_2)\left(n+\frac{2}{\beta}-1\right)+(m_1-m_2)^2\right]$} \\
 \textcolor{black}{A random density matrix ($\rho$) and a fixed density matrix ($\sigma$)}  & \textcolor{black}{$\displaystyle D^2_{\rho,\sigma}=\tr \sigma^2+\frac{\beta(n+m+2/\beta-1)}{\beta n m+2}-\frac{2}{n}$}\\
 \textcolor{black}{Two random density matrices ($\rho_1,\rho_2$)} &  \textcolor{black}{$\displaystyle D^2_{\rho_1,\rho_2}=\frac{\beta(n+m_1+2/\beta-1)}{\beta n m_1+2}+\frac{\beta(n+m_2+2/\beta-1)}{\beta n m_2+2}-\frac{2}{n}$}\\
 \vspace{-0.5cm}\\
\hline\hline
\end{tabular}}
\label{summary}
\end{table*}

\textcolor{black}{Distance measures other than Hilbert-Schmidt, such as trace distance and Bures distance, are acknowledged to be better suited for characterizations such as distinguishability of quantum states.} While large dimension asymptotic results exist for averages of these distances, it would be immensely useful if finite dimension results can be obtained. Moreover, it would be of interest to go beyond the mean of these distances and explore higher moments and distributions. Finally, one would also like to investigate the statistics of distances between random states distributed according to measures other than the Hilbert-Schmidt measure, such as Bures-Hall measure.

\acknowledgments
The author is grateful to Professor Karol \.{Z}yczkowski for fruitful correspondence. \textcolor{black}{He also thanks the anonymous referees for constructive comments.}

\appendix

\section{Relationship between normalization constants}

We prove here the relationship between the normalization constants $\cC$ and $C$ as given in Eq.~\eqref{CcC}. Since $\int d\rho \mathcal{P}(\rho)=1$, we obtain from Eq.~\eqref{Prho},
\begin{equation}
\cC^{-1}(t)=\int d[\rho] (\det \rho)^\alpha \delta(\tr\rho-t),
\end{equation}
where, as before, we have introduced the auxiliary variable $t$ inside the delta function. Taking the Laplace transform ($t\to s$), we obtain
\begin{equation}
\widetilde{\cC^{-1}}(s)=\int d[\rho] (\det \rho)^\alpha e^{-s\,\tr\rho}.
\end{equation}
We then consider the transformation, $\rho=(\frac{\beta}{2s}) W$ with $s>0$, so that $d[\rho]=(\frac{\beta}{2s})^{n[\beta(n-1)/2+1]} d[W]$. This gives
\begin{align}
\nonumber
\widetilde{\cC^{-1}}(s)&=\left(\frac{\beta}{2s}\right)^{\beta nm/2}\int d[W] (\det W)^\alpha e^{-\frac{\beta}{2}\,\tr W}\\
&=\left(\frac{\beta}{2s}\right)^{\beta nm/2} C^{-1}.
\end{align}
Taking the inverse Laplace transform we obtain
\begin{equation}
\cC^{-1}(t)=\frac{1}{\Gamma(\beta nm/2)}\left(\frac{\beta}{2}\right)^{\beta nm/2} t^{\beta nm/2-1}C^{-1}.
\end{equation}
Finally, setting $t=1$, we get
\begin{equation}
\cC^{-1}=\frac{1}{\Gamma(\beta nm/2)}\left(\frac{\beta}{2}\right)^{\beta nm/2} C^{-1},
\end{equation}
which yields the desired result appearing in Eq.~\eqref{CcC}.


\end{document}